%% file: main.tex
\documentclass[a4paper,final,12pt]{article} 
\usepackage[nohyperlinks]{acronym}
\usepackage[toc,page]{appendix}
\newcommand{\nocontentsline}[3]{}
\newcommand{\tocless}[2]{\bgroup\let\addcontentsline=\nocontentsline#1{#2}\egroup}

\usepackage[greek,english]{babel}

\usepackage{upgreek}
\usepackage[onehalfspacing]{setspace}
\usepackage{import}
\usepackage{float}
\usepackage[T1]{fontenc}
\usepackage[utf8]{inputenc}
\usepackage{verbatim} 
\usepackage{lipsum}
\usepackage{footnote}
\usepackage[stable]{footmisc}
\usepackage{enumerate}
\usepackage{parskip}
\usepackage{nameref}
\usepackage{fontawesome} 
\usepackage{pifont}
\usepackage{tcolorbox}
\usepackage[square,numbers,sort&compress]{natbib}

\usepackage{amsxtra,amssymb,amsthm,amstext} 

\usepackage{mathtools} 
\usepackage{bm}
\usepackage{chemformula}
\usepackage[detect-all,locale=US,exponent-product = \cdot]{siunitx} 
\DeclareSIUnit{\arbitraryunits}{\text{arb. units}}
\DeclareSIUnit\bar{bar}

\usepackage{graphicx} 
\usepackage[extendedchars]{grffile} 
\usepackage[position=below,tableposition=top,format=hang,labelfont=it,bf]{caption}
\usepackage{overpic}
\usepackage[labelfont=bf]{subcaption}
\usepackage{pdfpages}
\usepackage{picinpar} 
\usepackage{adjustbox} 
\usepackage{multirow}
\usepackage{tabularx}
\usepackage{longtable}
\usepackage{booktabs}
\usepackage{dcolumn}
\usepackage{array}

\usepackage[activate={true,nocompatibility},final,tracking=true,kerning=true,spacing=true,factor=1100,stretch=10,shrink=10]{microtype}  

\usepackage[final,
            pdfauthor={},
            pdfsubject={},
            pdffitwindow=true,
            pdftitle={Frequency locking in lasing ZnO nanowire pairs},
            bookmarks=true,
            bookmarksopen=true,
            bookmarksopenlevel=1,
            bookmarksnumbered=true,
            pdfborder={0 0 0}, 
            pagebackref=false
            ]{hyperref}
\usepackage{attachfile} 
\attachfilesetup{author={}, color=0 0 1}

\newlength{\mylength}
\AfterEndEnvironment{figure}{\noindent\ignorespaces}
\AfterEndEnvironment{table}{\noindent\ignorespaces}
\usepackage{fancyhdr}
\usepackage{xcolor}
\usepackage{colortbl}
\makeatletter
\renewcommand*{\@textcolor}[3]{
	\protect\leavevmode
	\begingroup
	\color#1{#2}#3
	\endgroup
}
\makeatother

\usepackage{fix-cm}

\usepackage{siunitx}
\usepackage[version=4]{mhchem}

\begin{document}
\frenchspacing
\setlength{\marginparsep}{2em}

\title{Frequency locking in lasing ZnO nanowire pairs}
\author{Ann-Kathrin Kollak$^{1,3,*}$, Lukas R. Jäger$^{1,*}$, Hark Hoe Tan$^2$, \\ Carsten Ronning$^{1,\dagger}$}
\date{}
\maketitle
$^{1}$Institute of Solid State Physics, University of Jena, Max-Wien-Platz 1, 07743 Jena, Germany\\
$^{2}$ARC Centre of Excellence for Transformative Meta-Optical Systems, Department of Electronic Materials Engineering, Research School of Physics, The Australian National University, Canberra, ACT 2600, Australia\\
$^{3}$Department of Physics, Paderborn University, Warburger Str. 100, 33098 Paderborn, Germany\\

\vspace{-0.5cm}
$^{*}$These authors contributed equally to this work.\\
$^{\dagger}$Corresponding author: carsten.ronning@uni-jena.de
\input{1_Paper}
\clearpage
\bibliography{main}{}
\bibliographystyle{achemso}
\clearpage
\input{2_SI}
\end{document}

%% file: 1_Paper.tex
\section*{Abstract}
Frequency locking between coupled laser systems provides a powerful mechanism for stabilizing and controlling coherent emission, yet its implementation and applicability down to the nanoscale remains unknown and unexplored. 
Here, we demonstrate optical coupling and frequency locking in closely spaced ZnO nanowire lasers operating in the extreme near field (gap < \SI{10}{\nano\meter}). We observe both full and partial frequency locking, manifested as the alignment of all or a subset of the lasing modes, by spatially controlling the optical excitation.
We also observe single-mode lasing in a coupled nanowire pair where the multi-mode lasing of individual nanowires is suppressed.
In contrast to previously reported coupled-cavity nanowire lasers, where spectral control arises from static filtering mechanisms such as the Vernier effect, our results indicate a dynamically established relationship between actively lasing nanowires. These findings establish frequency locking as a robust and tunable mechanism in nanowire lasers, opening new routes toward stabilized and controllable nanoscale light sources for integrated nanophotonic systems.

\section*{Introduction}
Frequency and phase locking between coupled oscillators are central mechanisms for stabilizing and controlling laser emission \cite{Adler1946,Lang1980}. 
In photonics, such locking and synchronization phenomena enable linewidth narrowing, improved coherence, and precise spectral control, and have therefore become key tools in applications ranging from high-power laser stabilization to optical communications \cite{Otsubo2017,Strogatz2018}. 
Optical injection and mutual coupling schemes have been widely explored in macroscopic and semiconductor laser systems, where they provide robust routes to frequency and phase locking over well-defined parameter ranges \cite{Buczek1973,Kobayashi1981,Wieczorek2005,Otsubo2017}.

Transferring these concepts to nanoscale light sources is of growing importance as photonic technologies move toward dense on-chip integration \cite{Miller2010}. Nanoscale lasers offer highly compact device footprints and can operate at low lasing thresholds \cite{Strauf2011}, making them promising building blocks for integrated photonic systems. However, their small mode volumes and strong optical confinement enhance light–matter interaction and can increase the spontaneous-emission coupling factor into the lasing mode, which in high-$\beta$ nanolasers can lead to enhanced photon-number fluctuations and reduced coherence/spectral stability \cite{Strauf2011}. Establishing controlled coupling between nanolasers could therefore provide a powerful route to stabilize and engineer their emission properties, extending well-established concepts of laser locking and synchronization to the nanoscale.

Semiconductor nanowire lasers are particularly attractive candidates in this context. Their high crystal quality and strong optical gain support room-temperature lasing in Fabry--Pérot cavities \cite{Huang2001,Agarwal2006}. When brought into close proximity, nanowires exhibit significant evanescent field overlap \cite{Huang2007}, enabling efficient near-field coupling \cite{Guo2009}. Previous work has shown that coupling between nanowire laser cavities can strongly modify their emission spectra. In particular, coupled-cavity configurations have been used to achieve single-mode lasing through spectral filtering mechanisms often described in terms of the Vernier effect, where only overlapping resonances of mismatched cavities are selected \cite{Xiao2011,Xu2012,Gao2013}. More generally, these studies demonstrate that coupling can be exploited to engineer the modal structure of nanowire lasers.
However, in these approaches coupling primarily acts as a static spectral filter determined by cavity geometry, rather than as a dynamic interaction between actively lasing oscillators. In contrast, the regime of frequency locking---where two nanolasers emit the same lasing lines---has not been directly observed experimentally in nanowire lasers. Theoretical studies predict that such systems can exhibit rich nonlinear dynamics, including stable locking, partial locking, and transitions to more complex behavior depending on detuning, coupling strength, and pumping conditions \cite{Adams2019,Adams2022}.

Here, we experimentally demonstrate optical coupling and frequency locking in closely spaced ZnO nanowire lasers operating at room temperature. By positioning nanowires in the extreme near field (gap < 10 nm), we observe clear signatures of mutual interaction, including alignment of multimode lasing spectra. We identify regimes of both full and partial frequency locking, where either all or only a subset of modes spectrally align between the nanowires. Crucially, we show that the locking behavior can be actively controlled by spatially varying the optical pump, which modifies the relative excitation of the two nanowires. Furthermore, we observed single mode lasing in one nanowire pair and excitation conditions configuration. This enables controlled access to different coupling regimes, including full locking, partial locking, and the breakdown of locking.

\section*{Results}
\begin{figure}[ht]
    \centering
    \includegraphics{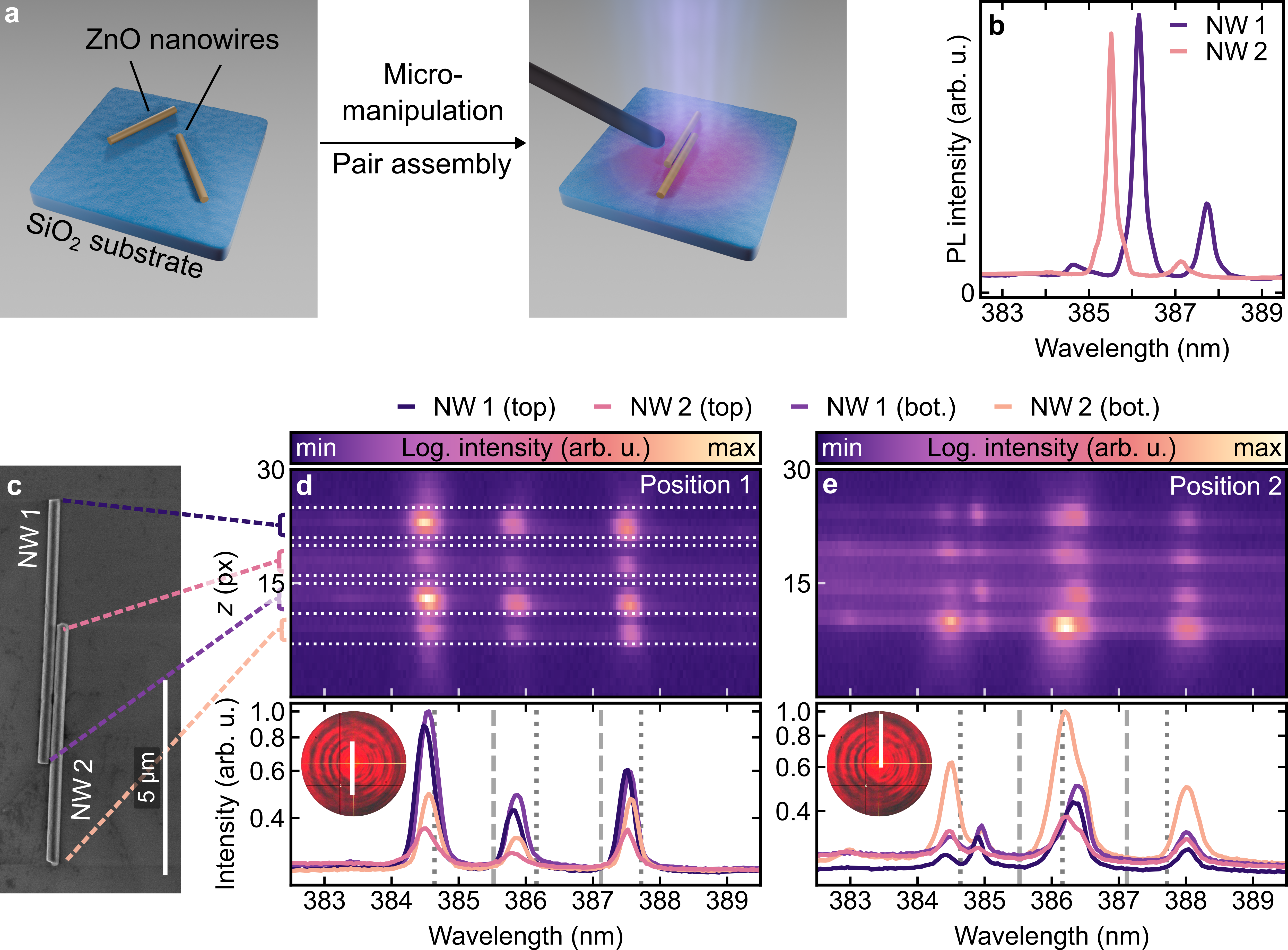}
    \caption{\footnotesize \textbf{a} Schematic depiction of the nanowire pair preparation. 
    \textbf{b} Lasing photoluminescence spectra of individual nanowires 1 and 2 (before pair assembly). 
    \textbf{c} SEM image of nanowire pair 1-2. Dashed lines indicate which emission track in the CCD image originates from which end facet.
    \textbf{d} Spatially resolved lasing spectra of nanowire pair 1-2 excited at position 1 in the inhomogeneous pump spot. Both NWs emit peaks at identical spectral positions. The gray lines mark the spectral positions of the lasing peaks emitted by NW 1 (dotted) and NW 2 (dashed) without coupling.
    The inset in the lower panel shows the camera image of the pump spot; the white line indicates the nanowire-pair--pump-spot configuration (position 1, see Figure \ref{fig:positions}\textbf{a} for larger image).
    \textbf{e} Spatially resolved lasing spectra at position 2.
    The inset in the lower panel shows the camera image of the pump spot; the white line indicates the nanowire pair position.
    }
    \label{fig:Pair1-2_locking}
\end{figure}

The nanowire (NW) pairs were assembled on a \ce{SiO2} on \ce{Si} substrate using a micromanipulator, as illustrated in Figure \ref{fig:Pair1-2_locking}\textbf{a}.
The pairs were assembled in a staggered fashion, giving rise to three sections along the nanowire axes. A middle coupling section, which facilitates energy transfer between the adjacent nanowires \cite{Huang2007}, and two protrusion sections, where each nanowire extends beyond the coupling section at either end of the nanowire pair.
Individual nanowires and nanowire pairs were pumped to the lasing regime using an enlarged laser spot that covered the nanowires fully.

\subsection*{Frequency Locking in Symmetric Nanowire Pairs}
First, we investigated two similar nanowires (NW 1, NW 2) with lengths of 6.76 and \SI{6.10}{\micro\meter} and diameters of 269 and \SI{315}{\nano\meter}, respectively (see Table \ref{tab:NW_geometry} for all nanowire geometric measurements). Their individual lasing spectra collected before pair assembly are presented in Figure \ref{fig:Pair1-2_locking}\textbf{b}. The spectra show the typical equidistant lasing peaks related to the longitudinal Fabry--P\'erot modes in the nanowire cavities. Each nanowire exhibits a characteristic mode spacing with peaks at different spectral positions. The spectra, in particular the free spectral range (FSR) of the individual nanowires are stable under different pumping conditions (see Section \ref{sec:SNW-stability}).

An SEM image of the two nanowires after pair assembly is shown in Figure \ref{fig:Pair1-2_locking}\textbf{c}. The lasing spectrum of this nanowire pair is presented in Figure \ref{fig:Pair1-2_locking}\textbf{d}. The spectrum was collected as a spatio--spectral intensity map, which is the main observable used to investigate optical coupling: the nanowire pair was positioned parallel with respect to the spectrometer slit (CCD-image-$z$-axis).
The emission was spectrally separated using a grating along the CCD-image-$\lambda$-axis direction.
Using this configuration we were able to resolve the emission from the quasi 1-dimensional system spatially and spectrally on the two-dimensional CCD simultaneously.

The spectral emission information of the four end facets of the two nanowires is thereby resolved individually, extracted by integration between the horizontal dotted lines in Figure \ref{fig:Pair1-2_locking}\textbf{d}, and plotted in the lower sub-panel. To ensure a clean separability of the nanowire end-facet specific emission, the protrusion lengths were chosen to be a few micrometers in length. It is clearly observable that nanowire pair 1-2 exhibits end-facet specific spectra with Fabry--P\'erot modes at identical spectral positions in contrast to their individual free-running mode positions.

This emission with identical FSR and Fabry--P\'erot peak positions through all end facets clearly indicates frequency locking as a result of evanescent coupling present in the overlapping section of the nanowire pair. The FSR and cavity length are related: $\Delta\lambda=\frac{\lambda^2}{2L\,n_g}$, where $n_g$ is the group index, and $L$ the cavity length \cite{Rabus2007,Saleh1991}. We extracted the FSR of each pair of modes, calculated a group index for each and thereby obtained an effective cavity length $L_\text{eff}$  (see Section \ref{sec:ng-model}), resulting in $L_\text{eff}\approx{\SI{6.9(0.2)}{\micro\meter}}$. The nanowire pair acts as a mutually coupled system in which both lasers are influenced by the evanescent field of the other and the FSR falls in between the value of the two free running cavities \cite{Wille2004,Dubois2018}. The extracted cavity length is closer to the SEM-determined length of NW 1 rather than NW 2, leading to the conclusion that NW 1 takes on the dominant role as the brighter "master" laser in this mutually (bidirectionally) coupled laser system. For comparison, the peak positions of the individual nanowire lasers taken from \ref{fig:Pair1-2_locking}\textbf{b} are indicated as vertical gray lines in the lower sub-panel of Figure \ref{fig:Pair1-2_locking}\textbf{d}. In that case, both individual nanowires act as detuned free running lasers with separate peak positions due to the 9\% difference in cavity length. This detuning lies withing the locking range of the system and the nanowire pair exhibited stable locking for a range of investigated pump powers from \SIrange{21.3}{48.6}{\kilo\watt\per\centi\meter\squared}, which corresponds to values up to 2.3 times the lasing threshold (see Section \ref{sec:lasing-threshold}). The large locking range is assumed to be a result of the low quality factors ($Q$) of the nanowire cavities ($\sim 500$ \cite{Maslov2003,Li2005,Wang2006,Ding2007}) and interaction in the strong coupling regime due the to extremely close coupling gap geometry, both of which broaden the locking bandwidth \cite{Kreinberg2019, Liu2020}.

The higher emission intensity of NW 1 and its role as the "master" laser in the frequency-locked system are consistent with it being pumped with a higher excitation intensity compared to NW 2 \cite{Kreinberg2019}. These asymmetrical pumping conditions were achieved by using an inhomogeneous pump laser spot and placing the pair at "position 1" within that spot. A camera image of the pump spot serves as a reference of the spatial intensity distribution incident on the sample surface and the nanowire pair (see inset in Figure \ref{fig:Pair1-2_locking}\textbf{d}, Figure \ref{fig:positions}\textbf{a}). We measured a  photoluminescence (PL) intensity map of the pump spot (see Figure \ref{fig:Laserspot}) to show that the camera image intensity reflects well the local excitation intensity.

\subsection*{Coexistence of Locked and Unlocked Modes}
Figure \ref{fig:Pair1-2_locking}\textbf{e} shows the the spatio--spectral intensity map for an excitation intensity of 2.0 times the lasing threshold (\SI{92.6}{\kilo\watt\per\centi\meter\squared}) and the extracted end-facet specific spectra when the nanowire pair is placed at position 2 of the pump spot (see inset in Figure \ref{fig:Pair1-2_locking}\textbf{e} and Figure \ref{fig:positions}\textbf{a}). Thus, NW 2 is now positioned in the intensity maximum and pumped with a higher intensity. The lasing spectrum shows a partial breakdown of the frequency-locked state that results in the Fabry--P\'erot modes splitting to reestablish the free running FSR of the individual nanowires. This splitting occurs on the blue side of the lasing spectrum. In contrast, the peak on the red side of the spectrum at \SI{388}{\nano\meter} remains frequency locked due to reduced mode confinement and the resulting stronger coupling efficiency at longer wavelengths \cite{Chen2017xxx}.

The cavity of origin of each mode can be inferred from the brightness of the lasing lines in the individual nanowires. We extracted effective cavity lengths of \SI{6.8(0.2)}{\micro\meter} and \SI{5.9(0.1)}{\micro\meter} for the lasing lines assigned to NW 1 and NW 2, respectively, in good agreement with the lengths of the two nanowires. However, these lasing lines are also outcoupled and amplified in the respective other wire due to the mutual coupling, causing them to appear in both spectra with different intensities. The reason for the partial breakdown of the frequency-locked state lies in both the pumping conditions and the geometry of the nanowires. The latter is connected to the $Q$-factor of the individual sub-cavities: Shorter nanowires such as NW 2 exhibit a lower $Q$-factor compared to longer nanowires as the result of the end facets acting as the dominant source of loss \cite{Arafin2013}. This increases the range in which they lock to another laser \cite{Liu2020}. Additionally, NW 1 exhibits an enhanced coupling efficiency toward NW 2 given its smaller diameter and the resulting reduced mode confinement (i.e., a stronger evanescent field). NW 1 is therefore better suited to act as the dominant laser ("master") in the system, leading to stable frequency locking at position 1. This frequency locking can be (partially) lifted when the pumping conditions favor NW 2 as the "master" laser of the coupled system. The persistence of one frequency-locked peak can be attributed to the small detuning of the individual cavities and the high locking range caused by the overall low $Q$-factors. These results underline that an active dynamic control over the lasing spectrum of the evanescently coupled nanowire pairs, in particular the switching on and off of a fully frequency-locked state, is possible. The results were reproduced with a similar nanowire pair (see Figure \ref{fig:Lasing_Pair_3-4_SI}).

\begin{figure}[ht]
    \centering
    \includegraphics{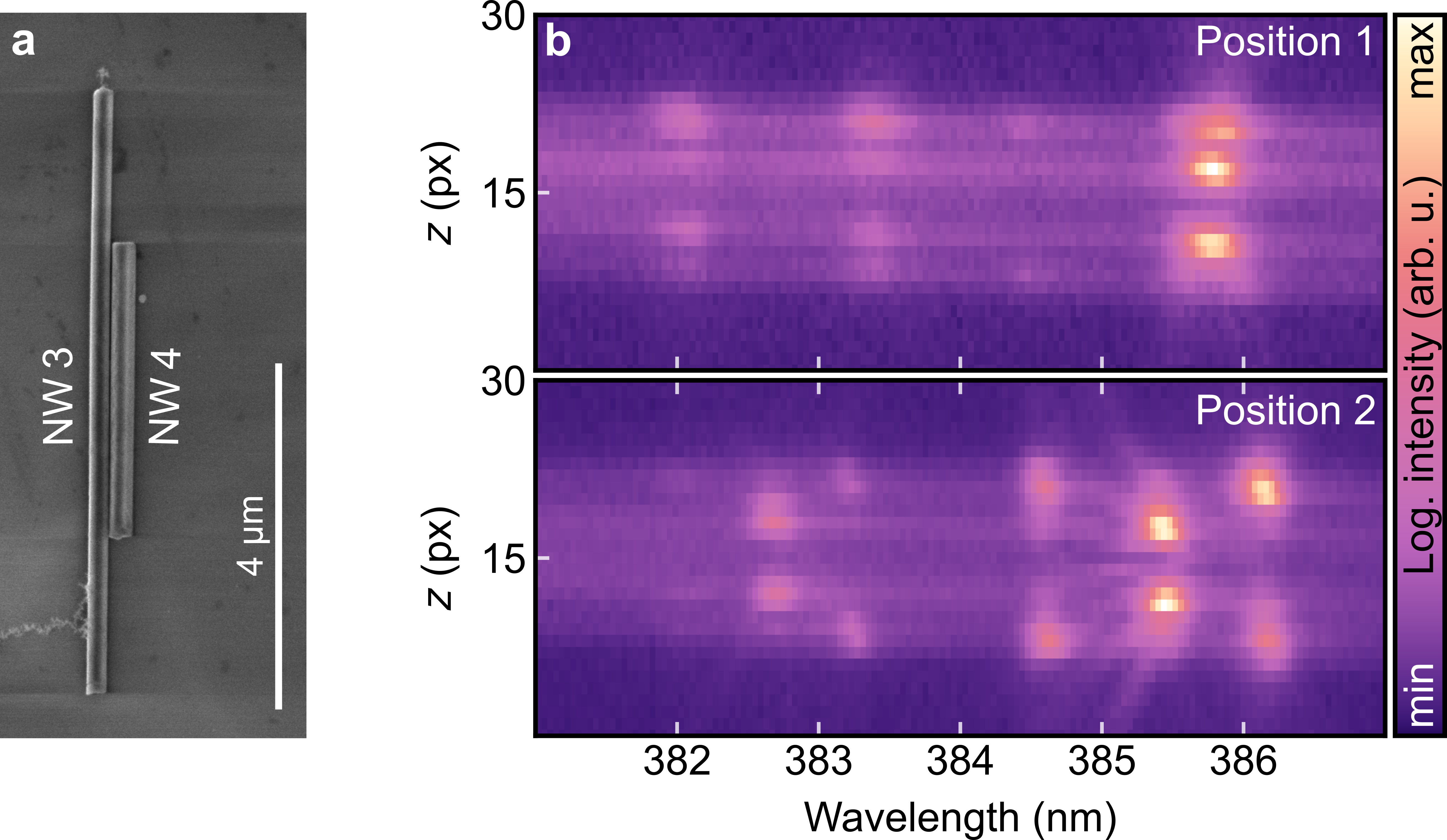}
    \caption{\footnotesize  \textbf{a} SEM image of nanowire pair 3-4. \textbf{b} Spatially resolved lasing spectra of nanowire pair 3-4 for two different positions in the inhomogeneous pump spot. Frequency locking can be observed for position 1. For position 2, the frequency locking breaks down and the nanowires emit spectra with an individual mode spacing that fits their respective cavity length. \normalsize}
    \label{fig:Pair5-6_locking_breakdown}
\end{figure}

\subsection*{On–Off Control of Frequency Locking}
We assembled a pair consisting of nanowires with a significant difference in length (see Table \ref{tab:NW_geometry}), to investigate the locking range and the influence of detuning between the coupled nanowires. The lengths were \SI{6.94}{\micro\meter} and \SI{3.42}{\micro\meter} for NW 3 and NW 4, respectively. Both ends of NW 3 protrude beyond the shorter nanowire (NW 4) as can be seen in the SEM image in Figure \ref{fig:Pair5-6_locking_breakdown}\textbf{a}. It is important to note that in this pair the coupling region extends over the entire length of NW 4. Two spatio--spectral intensity maps for two different pumping conditions are displayed in Figure \ref{fig:Pair5-6_locking_breakdown}\textbf{b}. They can be interpreted analogous to the results in Figure \ref{fig:Pair1-2_locking}. When NW 3 was excited with a higher intensity and can therefore be assumed to act as the dominant laser (position 1, see Figure \ref{fig:positions}\textbf{b}), frequency locking with one dominant laser line on the red side of the spectrum is clearly observed. This behavior persisted for excitation intensities of \SIrange{82}{148}{\kilo\watt\per\centi\meter\squared}. We extracted an effective cavity length of \SI{7.47(0.11)}{\micro\meter}, which is longer than NW 3, the longest individual cavity in the coupled system. A logical reason is an increase of the effective cavity length due to the mode oscillating between the two sub-cavities/nanowires as a result of the high coupling strength. The high coupling strength also leads to frequency pulling \cite{Fader1985} between the two significantly detuned cavities, resulting in a FSR that does not monotonically increase with the wavelength as would be expected for a Fabry--P\'erot cavity (see Equation \eqref{eq:mode-spacing}). Both nanowires adjust their emission spectra to accommodate a common FSR, with the peak on the red side of the lasing spectrum, which exhibits the highest coupling efficiency \cite{Chen2017xxx}, acting as an anchor. Given the significantly larger free-running FSR of the shorter nanowire (NW 4), the peak at \SI{384.5}{\nano\meter} does not match its intrinsic mode spacing and is therefore non-resonant and largely suppressed. In contrast, peaks located further from the dominant emission peak are still supported by the coupled system despite the significant detuning. The overall reduced emission through the lower end facet of NW 3 could be a result of a higher reflectivity of that end facet.

The lower sub-panel of Figure \ref{fig:Pair5-6_locking_breakdown}\textbf{a} shows the behavior of the same nanowire pair 3-4 when moved to position 2 in the pump spot. In this position, both nanowires were excited without a significant inhomogeneity (see Figure \ref{fig:positions}\textbf{b}). The spatio--spectral intensity map reveals that the nanowires do not share common lasing lines, a behavior that was observed for the investigated pump intensities of \SIrange{162}{225}{\kilo\watt\per\centi\meter\squared} (up to 1.4 times the lasing theshold). Consistent with the previous case of nanowire pair 1-2, the frequency locking breaks down if the shorter nanowire with the lower $Q$-factor is favored by the pumping conditions and excited with a higher excitation intensity.
Due to the substantial detuning a total breakdown is present in pair 3-4, including the strongly coupled modes on the red side of the lasing spectrum. In consequence, both nanowires act as individual free-running lasers and adopt a FSR that converts into an effective cavity length of \SI{6.84(0.11)}{\micro\meter} and \SI{3.39(0.11)}{\micro\meter}, consistent with their SEM-measured lengths of \SI{6.94}{\micro\meter} and \SI{3.42}{\micro\meter}, respectively. Overall, these results further demonstrate the interplay of geometry and the pumping conditions in the mutual coupling scheme present in these evanescently coupled nanowire laser pairs. They show that the frequency locking can be switched on and off completely when changing the relative excitation intensities of each cavity, if the nanowires are strongly detuned.

\subsection*{Absorption-Induced Single-Mode Lasing in Coupled Nanowires}
\begin{figure}[ht]
    \centering
    \includegraphics{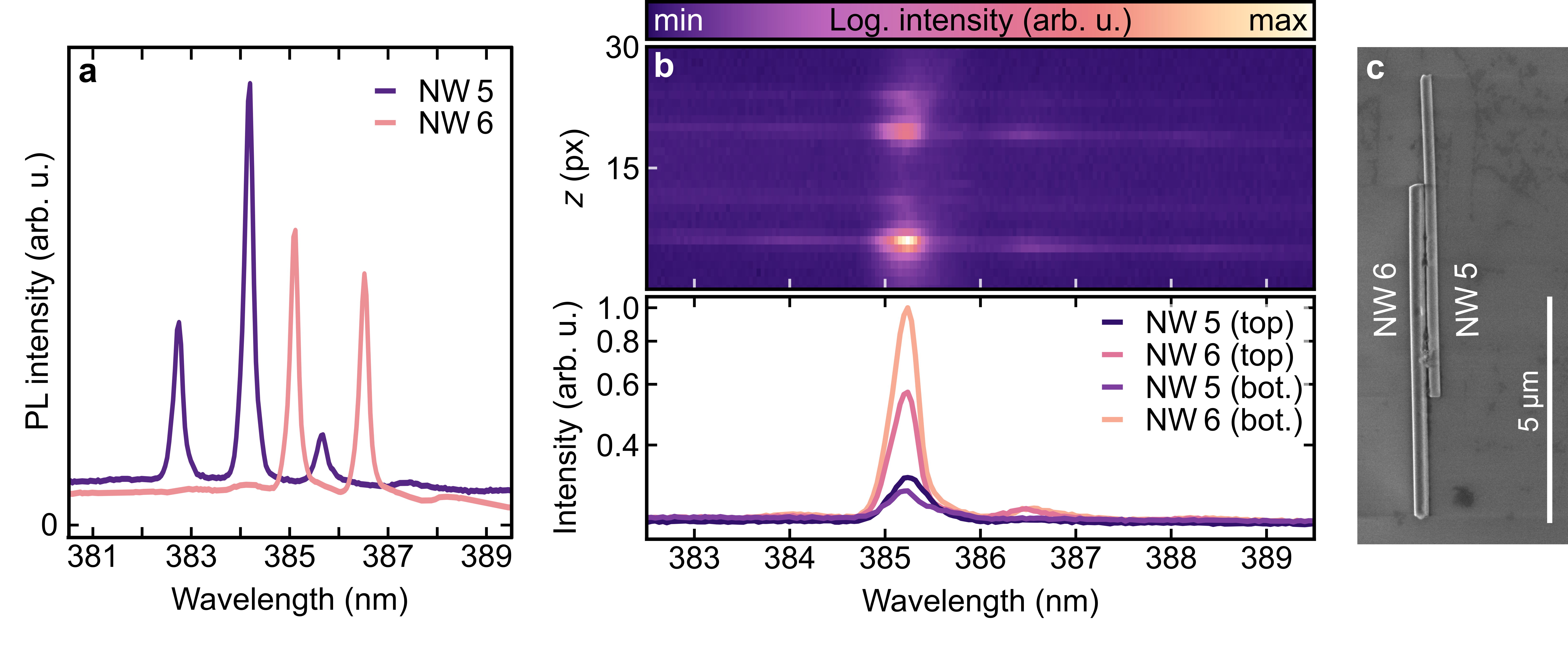}
    \caption{\footnotesize \textbf{a} Lasing photoluminescence spectra of nanowires 5 and 6 before pair assembly. \textbf{b} Spatially resolved lasing spectra of nanowire pair 5-6 at position 1 in the pump spot. \textbf{c} SEM image of nanowire pair 5-6. \normalsize}
    \label{fig:pair3-4_single_mode}
\end{figure}

Figure \ref{fig:pair3-4_single_mode} presents the results obtained for a third nanowire pair (NW 5 and NW 6) composed of nanowires with similar lengths: \SI{7.08}{\micro\meter} and \SI{7.36}{\micro\meter}. Their lasing spectra were collected before assembling the nanowire pair and are displayed in Figure \ref{fig:pair3-4_single_mode}\textbf{a}. Both individual nanowires exhibit a multimode lasing behavior with no spectral overlap of the mode positions. The spatio-–spectral intensity map recorded after pair assembly and the extracted end facet spectra shown in Figure \ref{fig:pair3-4_single_mode}\textbf{b} indicate that the pair, depicted in the SEM image in Figure \ref{fig:pair3-4_single_mode}\textbf{c}, exhibits single-mode lasing with a side-mode suppression ratio of approximately 17. A dominant lasing mode between 385 and \SI{386}{\nano\meter} is brightly emitted through the end facets of NW 6, which is excited with an excitation intensity of \SIrange{83}{140}{\kilo\watt\per\centi\meter\squared}, while NW 5 is placed in an excitation minimum. A camera image of the inhomogeneous pumping conditions is shown in Figure \ref{fig:positions}\textbf{c}).

A phenomenon that has been reported in the literature for coupled nanowire lasers is single mode lasing based on the Vernier effect, a static filtering mechanism that suppresses all but the spectrally overlapping lasing modes \cite{Xu2012, Gao2013, Xiao2011}. Such a mechanism is not expected to occur with the much shorter nanowires investigated in this work given the presumably lower $Q$-factors and the resulting dynamic behavior such as changes in the spectral mode positions due to frequency pulling of the two coupled lasers. Such a behavior resulted in (partial) frequency locking in the case of NW 1 and NW 2 with similar lengths, an effect that was reproducible with nanowire pair 5-6 (see Figure \ref{fig:Lasing_Pair_3-4_SI}). Therefore, we conclude that the single-mode lasing observed in this case cannot be explained by the Vernier effect but is an absorption based effect. Modes on the blue side of the lasing spectrum are suppressed by excitonic absorption present in areas of the nanowires that do not reach population inversion caused by the inhomogeneous pumping conditions \cite{Wille2016}. Modes on the red side of the lasing spectrum are strongly coupled to the non-excited NW 5 given the lower confinement \cite{Chen2017xxx} and are absorbed and suppressed. Consequently, only modes within a limited bandwidth of the lasing spectrum are sustained by the coupled system, resulting in single-mode lasing without the necessity of precisely matching two laser cavities.

\section*{Discussion}

We have demonstrated that ZnO nanowire laser pairs support a range of coupling regimes, including full frequency locking, partial locking, and the absence of locking, and that transitions between these states can be controlled through the spatial distribution of the optical pump. These observations establish evanescently coupled nanowires as a versatile platform in which both the strength and the symmetry of the interaction can be tuned in situ, without requiring structural modification of the cavities.

A central result of this work is that the observed behavior is governed by dynamic mutual interaction between actively lasing nanowires, rather than by static spectral filtering. In previously reported coupled nanowire laser systems, coupling has primarily been exploited to reshape the emission spectrum through coupled-cavity effects, often described in terms of the Vernier effect \cite{Xiao2011,Xu2012,Gao2013}. In such cases, the lasing modes are selected by the geometrical overlap of cavity resonances. In contrast, the alignment of multimode spectra observed here, including the emergence of a common FSR, indicates that the lasing states are determined by nonlinear interaction between the cavities. This behavior is characteristic of frequency pulling and locking in mutually coupled oscillators, where the emission frequencies shift toward each other and form collective modes of the coupled system.
The observed frequency locking demonstrates the emergence of a stable dynamical relationship between the nanowires.

The robustness of the observed locking further distinguishes this regime from previously studied coupled nanowire lasers. Frequency locking was observed over a wide range of excitation powers and for cavities with significant intrinsic detuning, including length differences on the order of several percent. This comparatively large locking range can be attributed to the moderate cavity quality factors of the nanowires, which are limited by end-facet losses \cite{Arafin2013}. The resulting broader linewidths reduce spectral selectivity and increase the tolerance of the system to detuning, thereby enabling efficient coupling even in geometries that would not support locking in high-$Q$ cavities. Together with the extremely close coupling gap arrangement of the nanowires, which enhances the evanescent field overlap, this places the system in a regime of strong coupling with a comparatively large locking range.

A key feature of the system is the ability to control the coupling dynamics through the excitation conditions. By varying the relative pumping of the two nanowires, we modify the balance of the interaction and thereby access different locking regimes, including transitions between locked and unlocked states. More generally, the coupled system exhibits an effective asymmetry, arising from both the excitation conditions and intrinsic differences between the nanowires, such as geometry and associated losses. Under such conditions, one nanowire can assume a dominant, "master" role, while the other follows its emission, consistent with symmetry breaking in mutually coupled laser systems \cite{Wieczorek2005}. These results demonstrate that the coupling dynamics are not solely determined by geometry, but can be influenced and partially controlled through the excitation conditions, providing a direct handle to manipulate the collective emission properties.

The observation of partial frequency locking provides further insight into the underlying interaction. In this regime, only a subset of modes remains common, while others revert to their individual cavity resonances. This behavior can be attributed to the wavelength dependence of the coupling efficiency, which is governed by the spatial extent of the optical modes and their evanescent fields. Modes at longer wavelengths exhibit weaker confinement \cite{Chen2017xxx} and therefore stronger coupling, allowing them to remain locked even when shorter-wavelength modes decouple. Partial locking thus reflects a competition between intrinsic cavity detuning and mode-dependent coupling strength, and highlights the multimode nature of nanowire lasers as an additional degree of freedom in coupled systems.

In addition to these dynamical effects, we observe that coupling can be harnessed to achieve single-mode lasing under appropriate conditions. While single-mode emission in coupled nanowire lasers has previously been realized through static spectral filtering mechanisms \cite{Xiao2011,Xu2012,Gao2013}, the behavior observed here arises from a combination of dynamic coupling, asymmetric excitation, and loss in the non-lasing sections of the structure. This indicates that single-mode operation can be achieved without the need for precise cavity engineering, instead relying on the interplay between coupling and gain distribution.

The present system is based on optically pumped nanowires assembled in a staggered geometry, which provides flexibility but also limits the degree of control over cavity parameters and coupling strength. Future work could extend this approach to more controlled architectures, including lithographically defined arrangements or electrically pumped nanowire lasers, enabling integration into photonic circuits. The demonstrated robustness and tunability of the coupling further suggest that arrays of nanowire lasers could be used to realize networks of synchronized nanoscale oscillators. Such systems are of interest for applications in on-chip coherent light sources, reconfigurable photonic systems, and neuromorphic or oscillator-based computing, where collective dynamics and controllable interactions play a central role \cite{Brunner2025,Ji2025,Torrejon2017}.

Overall, our results show that frequency locking can be reliably achieved in nanowire lasers and that it provides a powerful and flexible mechanism to control their emission. This extends concepts of laser locking and coupled-oscillator dynamics to the nanoscale and opens new opportunities for dynamically tunable nanophotonic devices.

\section*{Methods}
\subsection*{ZnO nanowire growth}
ZnO nanowires were synthesized via a vapor–liquid–solid growth process in a horizontal three-zone tube furnace. A mixture of ZnO and graphite powders (1:1 molar ratio) was used as the source material and placed in the central hot zone of the furnace. Silicon substrates coated with a \SIrange{10}{15}{\nano\meter} Au film were positioned downstream in a temperature gradient region and served as growth substrates. During growth, the thin gold film dewets at elevated temperature to form nanoscale liquid droplets, which act as catalytic seeds. These droplets absorb Zn-containing vapor species and, following the vapor–liquid–solid (VLS) mechanism, become supersaturated, leading to precipitation of crystalline ZnO at the liquid–solid interface and unidirectional nanowire growth.

Prior to growth, the furnace was evacuated to approximately \SI{7}{\milli\bar} and heated to $\sim$1050\,$^\circ$C. During growth, an Ar/\ce{O2} gas mixture (\SI{10}{sccm} each) was introduced to transport vapor-phase species and maintain oxidizing conditions. Typical growth durations of $\sim$\SI{1}{\hour} yielded nanowires with lengths of \SIrange{5}{30}{\micro\meter} and diameters in the range of \SIrange{100}{500}{\nano\meter}.

\subsection*{Nanowire pair assembly}
Following growth, the NWs were transferred onto silicon substrates with a \SI{300}{\nano\meter} thermally grown \ce{SiO2} layer using a dry-imprint method. This solvent-free process enabled the isolation of individual nanowires while preserving their optical properties.

Nanowire pairs were assembled in situ under an optical microscope using a micromanipulator (Kleindiek Nanotechnik MM3A-EM) equipped with a tungsten probe tip (Picoprobe PT-14-6705-B; apex diameter $\sim\SI{200}{\nano\meter}$). The manipulator allows independent motion along three axes with nanometer-scale precision.

Prior to optical measurements, individual NWs were slightly repositioned to release residual strain, which was found to influence the lasing mode positions \cite{Zaunick2024}. For pair assembly, one nanowire was laterally displaced using the probe until close contact with a neighboring nanowire was established, forming a staggered geometry with a defined coupling region. Due to the in situ configuration within the optical setup, the inter-wire separation could not be quantified with nanometer precision; instead, effective coupling was ensured by observing correlated motion of both nanowires during manipulation.

To minimize mechanical damage, nanowires with diameters of approximately \SI{300}{\nano\meter} were preferentially used, as thinner nanowires exhibited increased flexibility and a higher risk of fracture during manipulation. All assembly steps were performed under ambient conditions within the optical setup. Scanning electron microscopy (SEM) imaging was conducted only after optical characterization to avoid carbon deposition, which can increase optical losses and modify lasing characteristics.
NW 1 to NW 6 were selected from the same growth batch.

\subsection*{Optical characterization}
Micro-photoluminescence (µ-PL) spectroscopy was used to investigate the optical properties of individual and coupled ZnO nanowires. The nanowires were optically excited using the third harmonic (\SI{350}{\nano\meter}) of a $Q$-switched Nd:YAG laser (pulse duration $\sim$\SI{7}{\nano\second}, repetition rate \SI{100}{\hertz}).

The excitation power was controlled using a combination of neutral-density filters and a continuously variable attenuator. A fraction of the beam was directed onto a calibrated silicon photodiode to monitor the excitation power during measurements. The relation between photodiode signal and power at the sample position was determined by prior calibration.

The excitation beam was focused onto the sample using a near-ultraviolet microscope objective (50×, $ \text{NA} = 0.42$). For lasing experiments, the beam was intentionally defocused to a spot diameter of approximately \SIrange{10}{20}{\micro\meter} to ensure excitation of the entire nanowire or nanowire pair.

The sample was mounted on a three-axis translation stage with piezoelectric actuators, enabling precise positioning and alignment of individual nanowires with respect to the excitation profile. Bright-field imaging was used to identify nanowires and monitor their position within the pump spot.

The emitted photoluminescence was collected through the same objective and directed either to a CMOS camera for imaging or to a spectrometer for spectral analysis. Spectra were acquired using a Czerny–Turner spectrometer equipped with interchangeable gratings and a nitrogen-cooled CCD detector. Long-pass filters were inserted into the detection path to suppress back-reflected pump light.

For spatially resolved measurements, the nanowire axis was aligned parallel to the entrance slit of the spectrometer, enabling simultaneous spatial and spectral resolution of the emission on the CCD. This configuration allowed extraction of end-facet-specific spectra and the construction of spatio-–spectral intensity maps used to analyze coupling effects.

\section*{Author Information}
\subsection*{Corresponding Authors}
\*E-mail: carsten.ronning@uni-jena.de

\subsection*{Authors and Affiliations}
\textbf{Department of Physics, Paderborn University, Paderborn, Germany}
Ann-Kathrin Kollak.

\textbf{Institute of Solid State Physics, Friedrich Schiller University, Jena, Thüringen, Germany}
Lukas R. Jäger, Carsten Ronning, Ann-Kathrin Kollak.

\textbf{ARC Centre of Excellence for Transformative Meta-Optical Systems, Department of Electronic Materials Engineering, Australian National University, ACT, Australia}
Hark Hoe Tan.

\subsection*{Author Contributions}
A.-K.K. and L.R.J. contributed equally to this work. A.-K.K. and L.R.J. planned the experiments. A.-K.K. carried out the pair assembly and PL measurements. A.-K.K. and L.R.J. performed data analysis and results discussions. The manuscript was written by A.-K.K. and L.R.J. jointly, with feedback from all authors.

\subsection*{Notes}
The authors declare no competing financial interest.

\section*{Acknowledgements}
We acknowledge funding from the Deutsche Physiker Gesellschaft (DFG) in the context of the IRTG 2675 "Meta-Active" project (project number 437527638). 

\section*{Data availability}
The data that support the plots within this article and other findings of this study are available upon reasonable request from the corresponding author.

%% file: 2_SI.tex
\section*{Supplementary Information for Frequency locking in lasing ZnO nanowire pairs}

\setcounter{section}{1}
\setcounter{figure}{0}
\setcounter{table}{0}
\renewcommand{\thefigure}{S\arabic{figure}}
\renewcommand{\thetable}{S\arabic{table}}
\renewcommand{\theequation}{S\arabic{equation}}
\renewcommand{\thesubsection}{S\arabic{subsection}}

\subsection{Nanowire pair positions in the pump spot and estimation of the lasing threshold}
\label{sec:positions}
\label{sec:lasing-threshold}

\begin{figure}[h]
    \centering
    \includegraphics{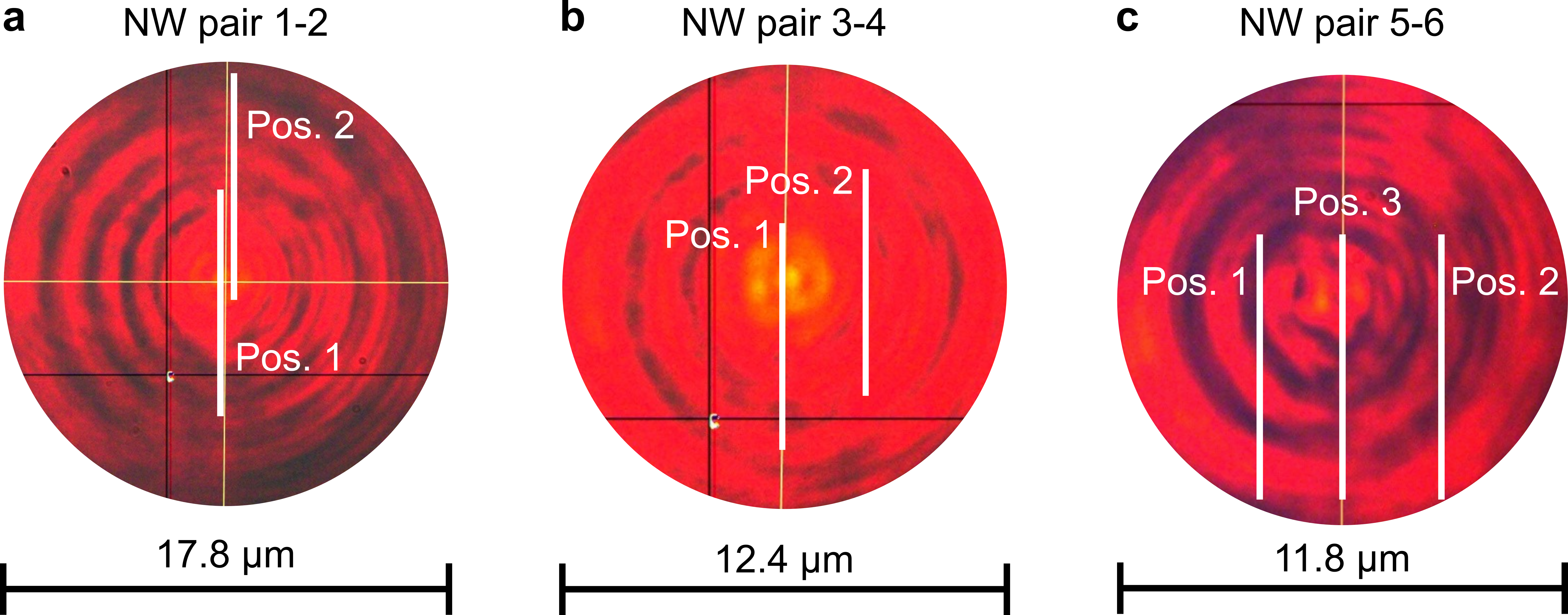}
    \caption{\footnotesize Camera images of the pump spots used to excite nanowire pairs \textbf{a} 1-2, \textbf{b} 3-4, and \textbf{c} 5-6. The white lines indicate measurement positions. \normalsize}
    \label{fig:positions}
\end{figure}

\label{sec:pump-spot}
\begin{figure}[ht]
    \centering
    \includegraphics{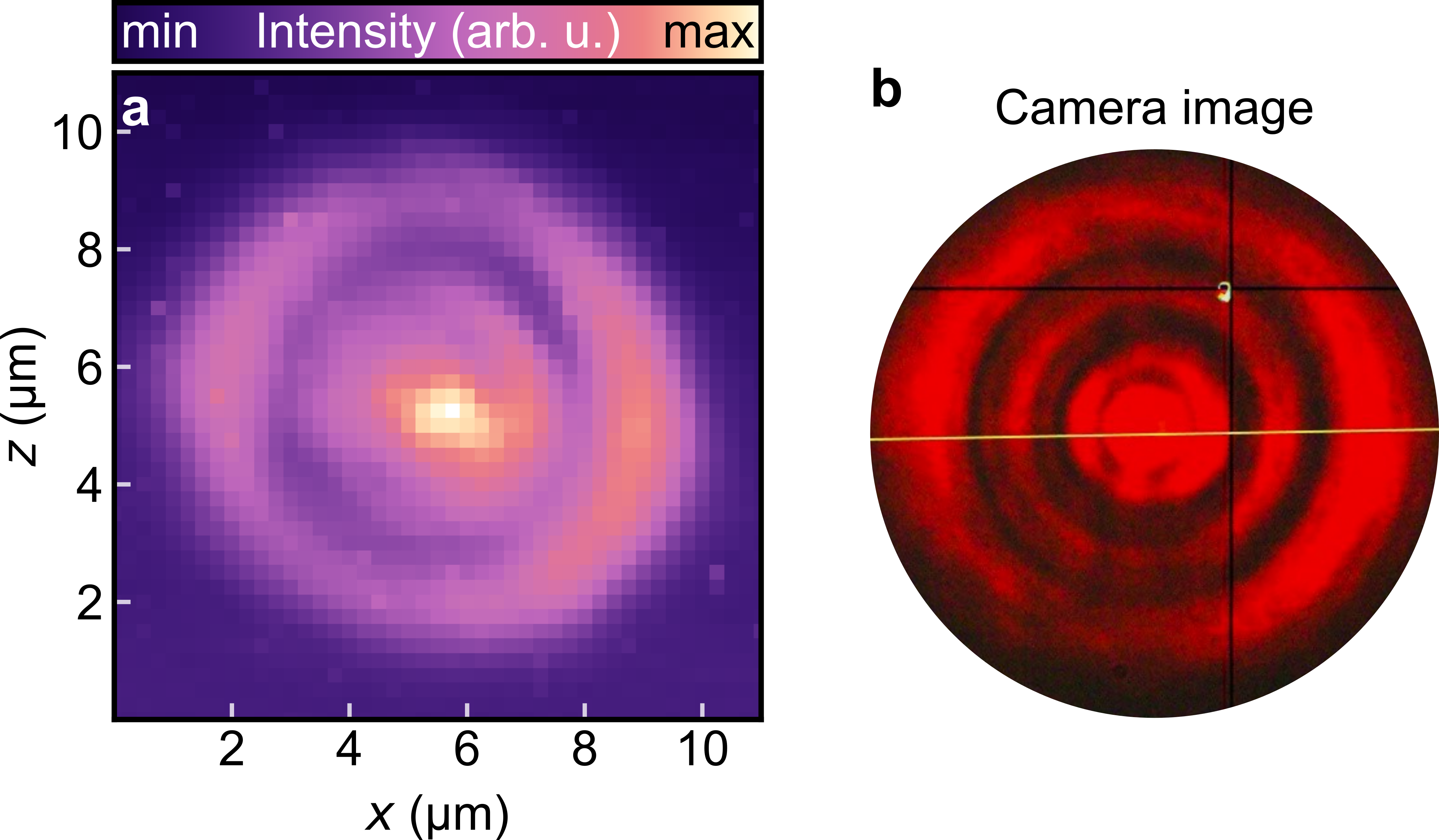}
    \caption{\footnotesize \textbf{a} PL intensity map recorded at a constant excitation power of \SI{102}{\kilo\watt\per\square\cm}, obtained by scanning the laser spot across a small ZnO particle. The camera image and the PL map exhibit similar intensity profiles. \textbf{b} Camera image of the broadened pump spot. The black and yellow lines are caused by dead pixels and alignment markers, respectively. \normalsize}
    \label{fig:Laserspot}
\end{figure}

In our experiments, the excitation is provided by an inhomogeneous pump spot (see Figure \ref{fig:positions}), resulting in a spatially varying intensity distribution. Consequently, absolute excitation power densities can only be specified as averages over the entire spot and do not accurately represent the local excitation experienced by a given nanowire or nanowire pair. This limitation becomes particularly relevant as individual nanowires (NW) and coupled pairs are deliberately positioned at different locations within the pump spot to exploit local intensity maxima and minima. Under these conditions, identical average excitation powers can correspond to substantially different effective local excitation levels, rendering direct comparison based on absolute values unreliable.

To enable meaningful comparison between measurements performed at different spatial positions, we therefore normalize the excitation to the lasing threshold of the respective nanowire system. Rather than extracting this threshold from conventional light–light (LL) curves, we adopt a more robust and reproducible criterion based on spectral signatures. In practice, LL curves obtained under our experimental conditions did not exhibit sufficiently well-defined slope changes to allow reliable fitting procedures, which we attribute to the combined effects of spatially varying excitation, limited sampling of excitation powers, and position-dependent coupling efficiencies.

Instead, the lasing threshold is operationally defined as the lowest excitation level at which well-resolved Fabry–Pérot modes emerge in the emission spectrum. Experimentally, this value is determined by recording emission spectra while incrementally increasing the excitation intensity from the low-power regime. The first occurrence of distinct, spectrally narrow modes is then taken as an estimate of the threshold. This procedure is repeated for each spatial configuration, allowing the definition of a relative excitation parameter 
$I_\text{meas} / I_\text{thresh}$, which enables comparison across different positions within the pump spot and between different measurements.

We note that this definition provides an estimate rather than an exact threshold value. Due to the discrete stepping of excitation power, the true threshold may lie between two measurement points. Moreover, the emergence of Fabry–Pérot modes can occur in the transition regime between amplified spontaneous emission and fully developed lasing. Nevertheless, the appearance of these modes constitutes a clear and experimentally accessible indicator of the onset of coherent feedback and modal gain. As such, it provides a consistent and physically meaningful reference point for comparing excitation conditions across all measurements presented in this work.

\subsection{Single nanowire Fabry--P\'erot mode stability}
\label{sec:SNW-stability}
\begin{figure}[ht]
    \centering
    \includegraphics[width=\textwidth]{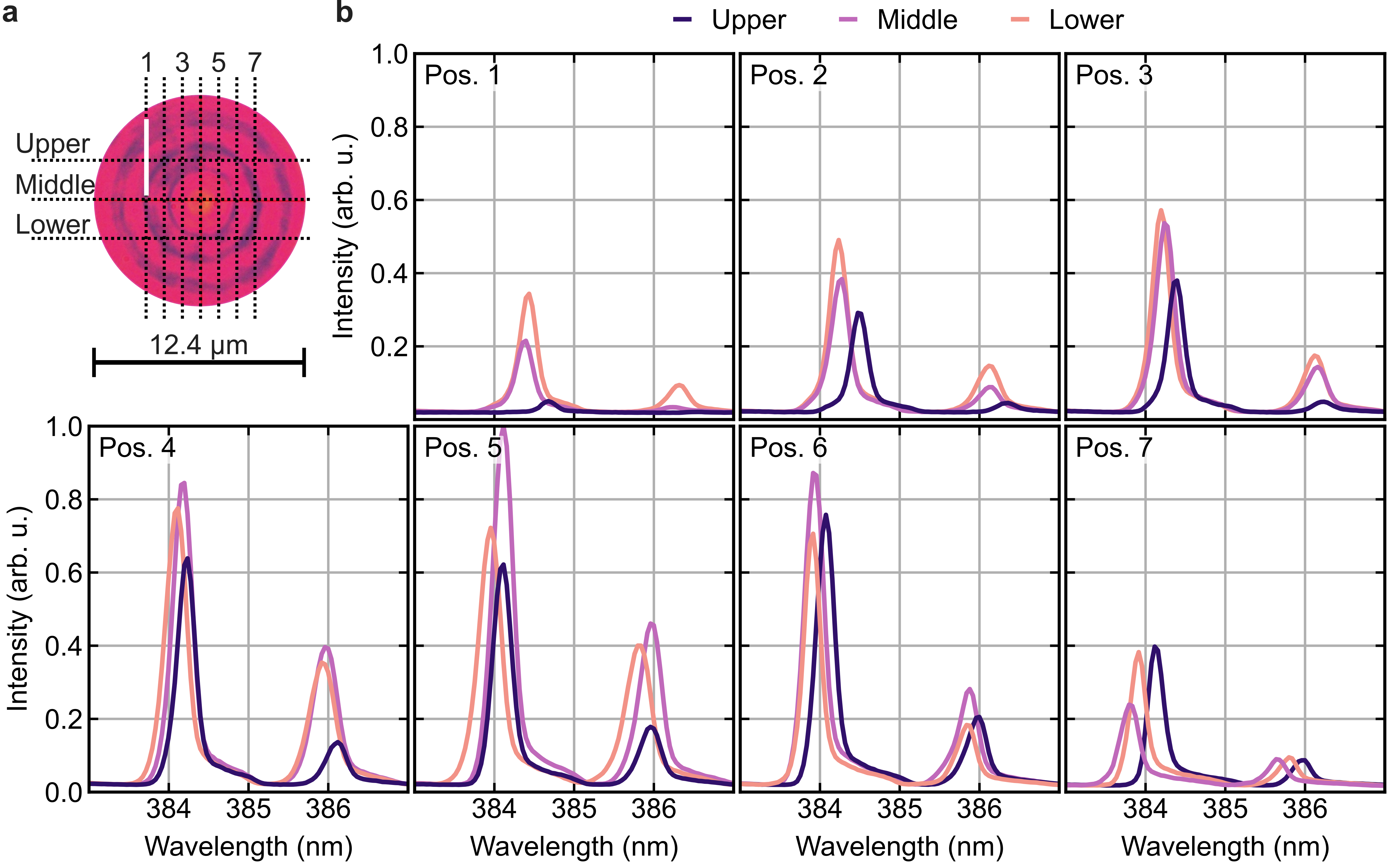}
    \caption{\footnotesize \textbf{a} Camera image of the excitation laser spot. Dotted lines mark 21 positions (7$\times$3) where lasing spectra were recorded. The white line indicates the nanowire at Pos. 1. \textbf{b} PL spectra of a single nanowire at the marked positions, recorded under a constant excitation intensity of \SI{233}{\kilo\watt\per\square\centi\meter}.\normalsize}
    \label{fig:SNW-stability}
\end{figure}
Figure \ref{fig:SNW-stability} presents lasing spectra recorded from a single nanowire (NW) at 21 distinct locations within the excitation spot. The corresponding spatial grid, along with a camera image of the pump spot, is shown in Figure \ref{fig:SNW-stability}\textbf{a}. Owing to the inhomogeneous intensity distribution of the spot, each location experiences a different effective excitation power. Consequently, the spectral profiles vary across positions, as illustrated in Figure \ref{fig:SNW-stability}\textbf{b}.
All measured spectra feature two prominent emission peaks. A gradual blueshift of these peaks is observed when moving from Pos. 1 to Pos. 7, which can be attributed to changes in the emission angle relative to the grating. The strongest emission intensities are detected at positions Pos. 4 to Pos. 6, corresponding to the center-right region of the excitation spot.
For most positions, the maximum emission intensity is observed at the middle height. Minor differences in peak wavelengths between the upper, middle, and lower measurement points are evident. These variations are smallest near the center of the spot (on the order of \SI{0.1}{\nano\meter}) and become more pronounced toward the edges, reaching up to \SI{0.3}{\nano\meter}. This behavior can be explained by the circular shape of the excitation spot, which leads to stronger variations in both the effective excitation density and its spatial gradient within the nanowire at edge positions (notably Pos. 1 and Pos. 2).
Overall, despite local variations, the peak wavelengths remain nearly constant across the excitation spot.

\subsection{Free-spectral-range derived effective cavity length---linear group index model}
\label{sec:ng-model}
The free spectral range (FSR) of the lasing nanowire Fabry--Pérot modes is related to the nanowire length for single NWs by the following relation \cite{Rabus2007,Saleh1991} :
\begin{equation}
  \Delta\lambda
  = \frac{\lambda^2}{2L}
    \left(
      n - \lambda \frac{\mathrm{d}n}{\mathrm{d}\lambda}
    \right)^{-1}
  = \frac{\lambda^2}{2L\,n_g}.
  \label{eq:mode-spacing}
\end{equation}

\begin{figure}[htb]
    \centering
    \includegraphics[draft=false]{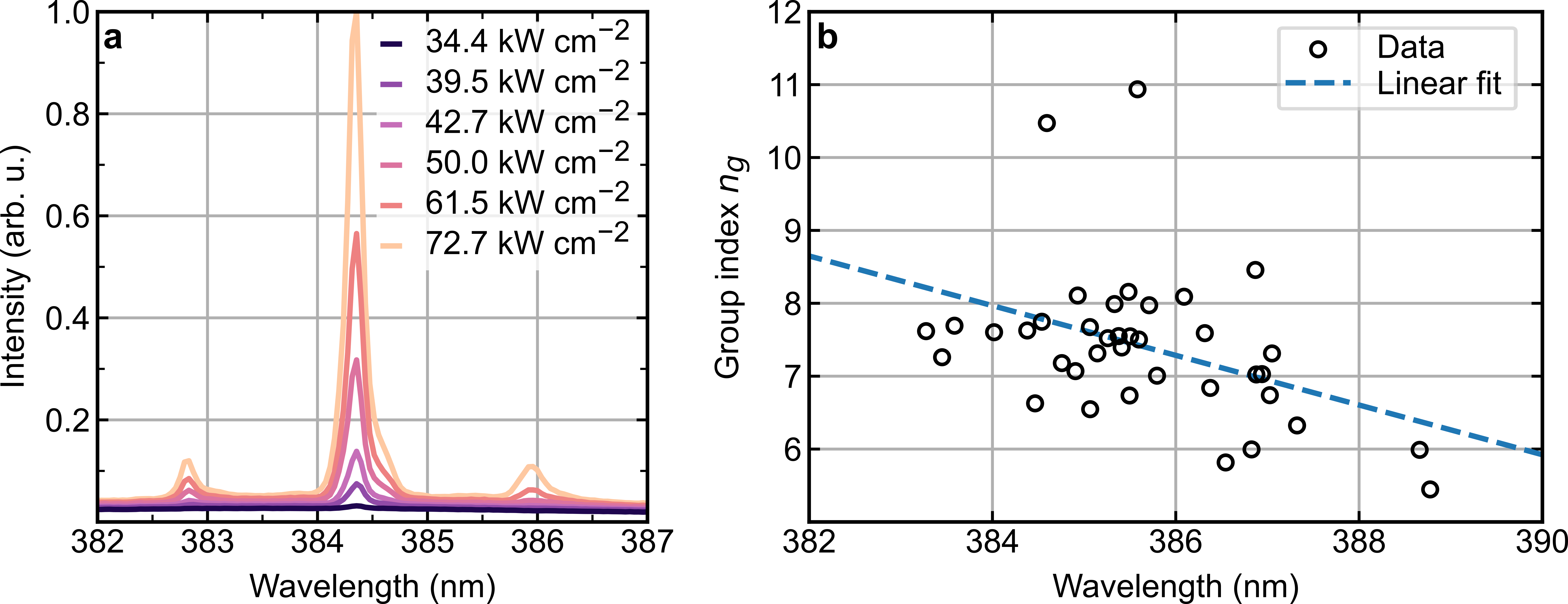} 
    \caption{\footnotesize \textit{Single nanowire lasing spectra and ensemble group index fit.} 
    \textbf{a} Lasing spectra of a representative ZnO nanowire at increasing pump power.
    \textbf{b} Group index $n_\mathrm{g}$ of the lasing mode (\textup{TE}\textsubscript{01}
 calculated from neighboring lasing lines of 21 single NWs (using Equation~\ref{eq:mode-spacing}) together with a subsequent linear fit. \normalsize}
    \label{fig:SI:ZnO:SNW-disp}
\end{figure}

Figure~\ref{fig:SI:ZnO:SNW-disp}a shows the lasing spectra of a representative single ZnO nanowire with increasing pump power. Three Fabry--Pérot modes are clearly visible and remain at constant wavelengths.
Using equation~\ref{eq:mode-spacing}, we can assign each pair of neighboring Fabry--Pérot modes a group index $n_g$. 
Figure~\ref{fig:SI:ZnO:SNW-disp}b shows this done for a collection of 21 different single NWs.
Each data point represents a pair of Fabry--Pérot modes.
The nanowire length $L$, required in equation~\ref{eq:mode-spacing}, was measured using a SEM.
We approximated the relationship with a linear fit (dashed blue line):
$n_\mathrm{g} = n_\mathrm{g}(0) + m\lambda$, yields $n_\mathrm{g}(0) = \num{138.8(48.2)}$ and $m = \SI{-0.341(125)}{\per\nano\meter}$.

\subsection*{Nanowire geometric measurements}
\label{sec:nw-geo}
\begin{table}[ht]
    \centering
    \caption{\footnotesize Geometry of NW pair 1-2, pair 3-4, and pair 5-6,, measured using the respective SEM images. The measurement uncertainties are \SI{15}{\nano\meter} (diameter), and \SI{30}{\nano\meter} (length). In NW Pair 5–6, the length of NW 6 defines the length of the overlap section and the significantly longer NW 5 extends beyond it on both sides, forming two overhanging sections.\normalsize}
    \label{tab:NW_geometry}
    \begin{tabular}{ccccc}
    \toprule
         & Diameter (nm) & Length (µm) & Protrusion (µm) & Overlap (µm)\\\midrule
        \textbf{NW1} & 296 & 6.76 & 3.12 & \multirow{2}{*}{3.55}\\
        \textbf{NW2} & 315 & 6.10 & 2.55 & \\\midrule
        \textbf{NW3} & 228 & 6.94 & 1.78, 1.76 & \multirow{2}{*}{3.42}\\
        \textbf{NW4} & 280 & 3.42 & -- & \\\midrule
        \textbf{NW5} & 255 & 7.08 & 2.40 & \multirow{2}{*}{4.68}\\
        \textbf{NW6} & 338 & 7.36 & 2.66 & \\\bottomrule
    \end{tabular}
\end{table}

\subsection*{Additional nanowire pair spatio--spectral intensity maps}
\label{sec:extras}
\begin{figure}[H]
    \centering
    \includegraphics{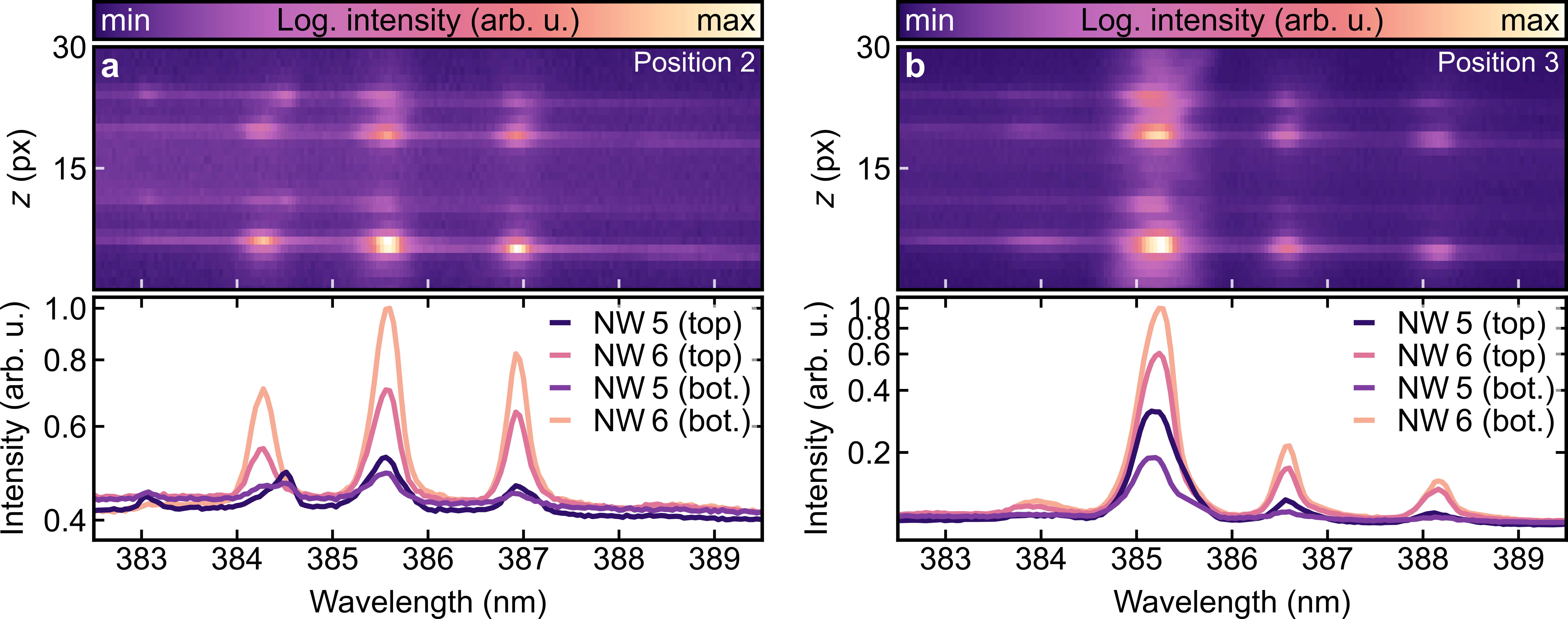}
    \caption{\footnotesize Spatio--spectral intensity map and end-facet specific spectra of NW pair 5--6 taken at \textbf{a} position 2 and \textbf{b} at position 3. The position 1 measurement is shown in Figure \ref{fig:pair3-4_single_mode}. At position 2 the pair exhibits partial frequency locking. At position 3 the pair exhibits full frequency locking. The corresponding pump spot for these measurements can be found in Figure \ref{fig:positions}. \normalsize}
    \label{fig:Lasing_Pair_3-4_SI}
\end{figure}